\documentclass[aps,prl,twocolumn,groupedaddress,superscriptaddress,showpacs]{revtex4}
\usepackage{graphicx}% Include figure files
\usepackage{bm}% bold math

\begin{document}
\title{Comment on "Superfluid turbulence from quantum Kelvin wave to classical Kolmogorov
cascades" by
J. Yepez, G. Vahala, L.Vahala and M. Soe
}
%%%%%%%%%%%%%%%%%%%%%%%%%%%%%%%%%%%%%%%%%%%%%%%%%%%%%%%%%%%%%%%%%%%%%
%%%%%%%%%%%%%%%%%%%%%%%%%%%%%%%%%%%%%%%%%%%%%%%%%%%%%%%%%%%%%%%%%%%%%

\author{Victor S. L'vov}
\affiliation{Department
of Chemical Physics, The Weizmann Institute of Science, Rehovot
76100, Israel}
\author{Sergey Nazarenko}
\affiliation{Mathematics Institute, Warwick University, Coventry, CV4 7AL, UK}

 
\maketitle

In the Letter~\cite{PRL}, the authors report on a high-resolution numerical simulation of turbulence governed by the Gross-Pitaevskii equation (GPE). The results are generally very impressive, in particular the obtained turbulent spectrum
scalings are quite convincing, and much cleaner than in   previous works.
However, their  interpretation of observed  scaling laws seems to us questionable:

\textbullet~Firstly, the authors observed
a $k^{-3}$ scaling for wavelengths  $\lambda$ less than the mean vortex core radius $a$, and  they  relate it to the  ``Kelvin waves". But in all previous literature on quantum turbulence, including the papers cited by the authors,  ``Kelvin waves" are understood as bending oscillations of the the quantized vortex lines    with $\lambda\gg a$.  Thus, any interpretation of observed 
$k^{-3}$ scaling should be based on   a physics which is completely different from that of the   Kelvin waves.
Namely, the long Kelvin waves (with $\lambda\gg a$) propagates \emph{along} the vortex lines and thus they have \emph{ one-dimensional} (1D) nature. Discussing   the \emph{short} waves (with $\lambda\ll a$)   one should account for the fact  that from  the very definition of the vortex core size $a$  in the range of scales $\lambda\ll a$ the nonlinear terms in GPE are much smaller  than the linear ones. Therefore the \emph{short} waves do  not ``feel" the vortex cores and propagate almost freely in the \emph{three-dimensional} (3D) space, with the mean-free path essentially exceedsing the wave-length $\lambda$.
In other words, in the considered case we are dealing with   a \emph{weakly nonlinear wave system}, and perhaps the weak turbulence theory~ \cite{ZLF,DNPZ}  should be employed to explain the scaling in this range. An important lesson of this theory is that the physics of 3D wave turbulence is completely different from that in 1D case. For example, the leading nonlinear interaction of 1D Kelvin waves is 6-wave scattering, while the interaction of considering here 3D waves should be dominated by 4-wave scattering. Moreover, in 3D case one can apply the wave kinetic equations~\cite{ZLF} for description of weak wave turbulence, while in 1D the correlations of the waves phases  are much stronger and the application of the kinetic equation is less obvious.
 %We are even not talking about that
  Besides, even simple dimensional arguments are very sensitive to the dimensionality of the space, etc.
 %And so on, and so forth.

 \textbullet~Secondly, the energy spectrum observed at the large  scales is close to  $k^{-5/3}$, and the authors   interpret it as the Kolmogorov spectrum. However the
 Kolmogorov phenomenological model  is only relevant to incompressible (or nearly incompressible) turbulence,
whereas, according to Figure 5 of the online supplement to this paper (and also the respective
movie) the compressible energy  is several times greater than the
incompressible one in this range of scales. It means that we are dealing with strongly compressible turbulence which should  be dominated by acoustic motions, namely by shocks or  random sound waves, and not by vortices as in the Richardson-Kolmogorov cascade. It therefore remains to be explained why a spectrum with exponent
close to Kolmogorov $-5/3$ is observed for the energy spectrum in the present work.
There are two reference theories that might help to resolve this mystery~\cite{ZLF}.
The first one, developed by  Zakharov and Sagdeev (ZS) \cite{ZS},
deals with turbulence of random weakly nonlinear acoustic waves
 and it predicted a spectrum
with exponent $-3/2$, a value very close to Kolmogorov $-5/3$ (the difference
is only $1/6$)
The second theory, developed by Kadomtsev and Petviashvill (KP) \cite{KP}, considers
turbulence dominated by acoustic shocks,
in which case the spectral exponent is $-2$. As we see, Kolmogorov $-5/3$ is
in between the ZS and the KP predictions for the acoustic dominated regimes,
and its observation could possibly be due to coexistence of strong shocks
and weakly nonlinear sound waves in the computed system.
Clearly, much work remains to be done for finding the character of the
turbulent motions (vortices vs shocks vs random sound waves), and one
should refrain from premature interpretations. % until such work is done.

 \textbullet~To summarize our comment: both the Kelvin wave and the Kolmogorov turbulence interpretations presented in  the Letter are misleading, and
  much more theoretical analysis needs to be done for the interpretation of the important numerical results obtained by the authors~\cite{PRL}.

\end{document}